\documentclass[prl, showpacs,floatfix,twocolumn,superscriptaddress]{revtex4}

\usepackage{latexsym}
\usepackage{amsmath}
\usepackage{amsfonts}
\usepackage{graphicx}
\usepackage{amssymb}
\usepackage{dsfont}
\usepackage{color}

\newcommand{\ket}[1]{\ensuremath{|#1\rangle}}
\newcommand{\bra}[1]{\ensuremath{\langle #1 |}}

\newcommand{\be}{\begin{equation}}
\newcommand{\ee}{\end{equation}}
\newcommand{\mc}[1]{\ensuremath{\mathcal{#1}}}
\newcommand{\bc}{\begin{center}}
\newcommand{\ec}{\end{center}}

\newcommand{\ve}{\varepsilon}

\newcommand{\vro}{\varrho}

\newcommand{\hf}{\mc{H}_{\text{FE}}}
\newcommand{\rd}{\vro_{\text{D}}}

\newcommand{\kn}{\ensuremath{K}}

\newcommand{\Si}[2]{S_{#1}^{(#2)}}

\begin{document}

\title{Dissipation induced Tonks-Girardeau gas of photons}

\author{M. \surname{Kiffner}}
\affiliation{Technische Universit\"at M\"unchen, Physik-Department I, 
James-Franck-Stra{\ss}e, 85748 Garching, Germany}

\author{M.~J. \surname{Hartmann}}
\affiliation{Technische Universit\"at M\"unchen, Physik-Department I, 
James-Franck-Stra{\ss}e, 85748 Garching, Germany}


\begin{abstract}
A scheme for the generation of a  Tonks-Girardeau (TG) gas of photons with   purely 
dissipative interaction   is described. 
 We put forward  a master equation approach for the description of stationary light in atomic 
four-level media  and 
show that, under suitable conditions, two particle decays are the dominant photon loss mechanism. 
These dissipative two-photon losses increase  the interaction strength
by at least one order of magnitude  as compared to dispersive two-photon processes
and can drive the photons into the TG regime. 
Our scheme allows for measurements of various characteristic correlations of the TG gas via
standard quantum optical  techniques, including quantities that
distinguish it from free fermions. 
\end{abstract}

\maketitle

Quantum mechanics categorizes particles into fermions or bosons. 
In three dimensions only these two categories are possible, whereas more exotic anyons 
can exist
in two dimensions~\cite{wilczek}. In one dimension, the particle statistics can not be 
considered without taking inter-particle interactions into
account~\cite{bloch:08}. 
A prominent example are bosons that interact via strong repulsive forces in a one-dimensional setting and can 
enter a Tonks-Girardeau (TG) gas regime~\cite{girardeau:60}, where they behave with 
respect to many observables as if they were fermions. A TG gas can be described as the
strong interaction limit of the Lieb-Liniger model~\cite{lieb:63}.

Strong correlations in many-particle systems, such as in the TG gas, give rise to 
interesting and partly not yet well understood physics.
A substantial amount of  research 
is thus currently devoted to these systems and 
progress in cooling and trapping of atoms and ions has opened up possibilities to 
study strongly interacting many-body systems experimentally
with unprecedented precision.  Eventually, this progress enabled  the 
observation of a TG gas of atoms in an optical lattice~\cite{paredes:04}.
Later, an experiment~\cite{syassen:08}  with cold molecules showed that not only 
elastic interactions, but even two-particle losses alone  are able to create 
a TG gas  where two molecules never occupy the same position, thereby 
avoiding dissipation of particles. 
This counterintuitive result can be regarded as a manifestation of  
the quantum Zeno effect~\cite{syassen:08}. 

In contrast to atoms, photons are massless particles that do not interact at all.
Nonetheless, effective many-body systems of photons and polaritons can be generated
by employing light matter interactions. This concept has been introduced
recently~\cite{HBP06,HBP08,GTCH06,ASB06} and is currently receiving increasing
attention~\cite{chang:08,GTI+09,HP07,RF07}.
To enter the strongly correlated regime and access its rich physics, 
sufficiently strong effective interactions are needed.
Suitable experimental setups, like arrays of coupled microcavities 
doped with  emitters~\cite{ADW+06} or optical fibers that couple to 
atoms~\cite{chang:08,bajcsy:09}, thus need to combine
strong photon-emitter coupling and low-loss photon propagation. 
In all these setups, the main challenge for realizing strong correlations is to make the 
polariton-polariton interactions much stronger than
photon losses  which are inevitably present in every experiment. 

%
\begin{figure}[b!]
\vspace*{-0.2cm}
\includegraphics[scale=0.9]{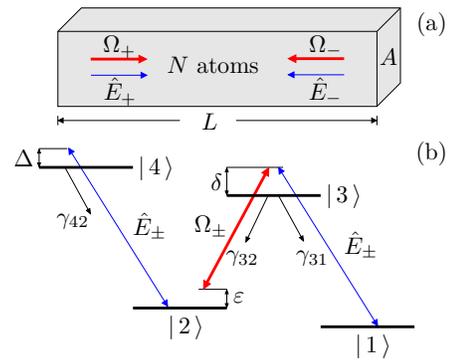}
\vspace*{-0.2cm}
\caption{\label{fig1} 
(Color online) (a) Considered setup of 
$N$ atoms   confined to an interaction volume of length $L$ and transverse area $A$. 
$\Omega_{\pm}$ are the Rabi frequencies of the classical control fields, and 
$\hat{E}_{\pm}$  are the quantum probe fields. 
(b) Atomic level scheme.   $\gamma_{ij}$ is the full decay rate 
on the $\ket{i}\leftrightarrow\ket{j}$ transition, 
$\delta$ and $\Delta$ label the detuning of the probe fields with states $\ket{3}$ and 
$\ket{4}$, respectively, and $\ve$ is the two-photon detuning. 
}

\end{figure}
Here we present an effective many-body system of polaritons 
where the ubiquitous but usually undesired dissipative 
processes become the essential ingredient for the 
creation of strong many-particle
correlations. 
This paradigm shift allows us to relax some conditions on the model parameters such that 
the achievable nonlinearities in our approach are at least an order of magnitude 
larger than their conservative counterparts~\cite{HBP06,HP07,chang:08}.
In particular, we show  that the dissipative nonlinearities in our system 
give rise to a TG gas of photons. 
{For this regime,  fermionic (e.g. Friedel oscillations) as well as
non-local (e.g. the single particle density matrix) correlations 
  of the TG gas can be measured via
standard quantum optical  techniques. 

We consider photons guided in an optical fiber that interact with 
nearby  atoms~\cite{chang:08,bajcsy:09}, 
where stationary light~\cite{bajcsy:03}
is created via Electromagnetically Induced Transparency 
(EIT)~\cite{fleischhauer:05}. 
As compared to coupled microcavities, the fiber approach is 
appealing  due to the low photon loss of the fiber and 
since the longitudinal trapping of light is done optically, thus avoiding the  
need to build many mutually resonant cavities.  We thus focus on this setup here. 
However, our mechanism for building up correlations works equally well in cavity arrays,   
and the dissipative nonlinearities we discuss here are always stronger than their 
conservative counterparts independent of the geometry of the experimental device~\cite{vinck:07}.

We start with a more detailed description of 
our one-dimensional model shown in Fig.~\ref{fig1}. 
Each of the $N$ atoms interacts with  control and probe  fields denoted by $\Omega_{\pm}$ and 
$\hat{E}_{\pm}$, respectively. 
The control fields of frequency $\omega_c$ are treated  classically 
and $\Omega_+$ ($\Omega_-$) labels  the Rabi frequency 
of the control field propagating in the positive (negative) $z$ direction.  
In addition, we assume that the control fields are spatially homogeneous but 
may depend on time. 
The probe fields $\hat{E}_+$ and $\hat{E}_-$ are quantum fields 
that propagate in the positive and negative $z$ direction, respectively. They  
are defined as $\hat{E}_{\pm}(z) = \sum_{\kn } a_{\pm \kn }e^{\pm i \kn  z}$, 
where $a_{\pm \kn}$ are photon annihilation operators. The wave numbers $\kn$  are  
positive and of the order of the wave number $k_c$ of the control field.

We model the  time evolution of the atoms and the quantized probe fields  by a 
master equation~\cite{breuer:os} for their density operator $\vro$, 
$\dot{\vro}= - \frac{i}{\hbar}[ H ,\vro] + \mc{L}_{\gamma}\vro$, 
where $\mc{L}_{\gamma}\vro$ describes spontaneous emission from states $\ket{3}$ and $\ket{4}$, 
and the full decay rate on the transition $\ket{i}\leftrightarrow\ket{j}$ is denoted by 
$\gamma_{ij}$ (see Fig.~\ref{fig1}). 
In a rotating frame that removes the time-dependence of the classical laser 
fields, the  system Hamiltonian $H$ reads 
$H  = H_0 + H_{\Lambda} + H_{\text{NL}}$, where 
\begin{align}
 H_0 = & -\hbar 
 \sum_{\kn }(\omega_p-\omega_{\kn })\left( a^{\dagger}_{\kn } a_{ \kn } 
+ a^{\dagger}_{- \kn } a_{ -\kn } \right) \notag \\
& -\hbar 
\sum_{\mu=1}^N\left[ \varepsilon A_{22}^{(\mu)} + \delta A_{33}^{(\mu)} + 
(\Delta + \varepsilon)A_{44}^{(\mu)}\right] 
\end{align}
describes the free time evolution of the atoms and the probe fields. 
$A_{ii}^{\mu} = \ket{i_{\mu}}\bra{i_{\mu}}$ is a projection operator onto 
state $\ket{i_{\mu}}$ of atom $\mu$, the energy of level $\ket{i}$ is 
$\hbar \omega_i$ (we set  $\omega_1=0$), and transition frequencies are denoted 
by $\omega_{ij}=\omega_i-\omega_j$. We denote the central frequency of the 
probe pulse  by $\omega_p$. The detuning 
of the probe field with respect to the transitions 
$\ket{3}\leftrightarrow\ket{1}$  and $\ket{4}\leftrightarrow\ket{2}$ 
is labeled by $\delta  = \omega_p - \omega_{31}$ and $\Delta =  \omega_p - \omega_{42}$, 
respectively, and $\ve  = (\omega_p-\omega_c) -\omega_2$  is the two-photon detuning. 
The interaction between the atoms and   the  probe and  control 
fields  is described by $H_{\Lambda} + H_{\text{NL}}$, with 
\begin{align}
H_{\Lambda} = & - \hbar\sum_{\mu=1}^{N}\Big\{ \Si{32}{\mu} 
\left[ \Omega_+(t) e^{i k_c z_{\mu}} + \Omega_-(t) e^{-i k_c z_{\mu}}\right] \notag \\
& \hspace*{0.5cm} + g_1 
\Si{31}{\mu}  \left[ \hat{E}_+(z_{\mu}) +\hat{E}_-(z_{\mu}) \right] \Big\}
 +\text{h.c.}   \,, \\
H_{\text{NL}} =  &  - \hbar g_2 \sum_{\mu=1}^{N}
 \Si{42}{\mu}  \left[ \hat{E}_+(z_{\mu}) +\hat{E}_-(z_{\mu}) \right]
 +\text{h.c.}  \,.
\end{align}
Transition operators of atom $\mu$ at position $z_{\mu}$ are defined as 
$\Si{ij}{\mu} = \ket{i_{\mu}}\bra{j_{\mu}}$ ($i\not=j$), 
and $g_1$ and $g_2$ are the single-photon Rabi frequencies on the 
$\ket{3}\leftrightarrow\ket{1}$ and $\ket{4}\leftrightarrow\ket{2}$ transitions, respectively. 
In the following, we assume that the Rabi frequencies of the control fields are  
identical (and real) and set $\Omega_+ = \Omega_- = \Omega_c$.   With this choice, 
the interaction   of the  probe and  control fields with 
the $\Lambda$-subsystem formed by states $\ket{1}$, $\ket{2}$ and $\ket{3}$ 
allows to store the probe field inside the medium~\cite{bajcsy:03}. 
On the other hand, the coupling of the probe fields to the 
$\ket{4}\leftrightarrow\ket{2}$ transition creates an effective 
photon-photon interaction~\cite{ISWD97}.  

Next we outline the approach we developed to reduce the master equation 
$\dot{\vro}= - \frac{i}{\hbar}[ H ,\vro] + \mc{L}_{\gamma}\vro$ for the atoms 
and quantized probe fields into a master equation solely for dark-state 
polaritons~\cite{fleischhauer:00}, formed by collective excitations of photons and atoms.
We represent the quantum state of dark-state polaritons 
by a density matrix $\rd$ comprised of dark-states
$ \ket{\alpha} =
 \prod_{k=1}^{N_{\alpha}}(1/\sqrt{n_k!})\big(\psi_k^{\dagger}\big)^{n_k}\ket{0}$ 
 that satisfy 
$H_{\Lambda} \ket{\alpha} = 0$,  
and the vacuum state 
$\ket{0}=\ket{\{0\}_{\text{phot}};1_1,\ldots,1_N}$ is the state where  
all photon modes of the probe fields are empty and all atoms are in state $\ket{1}$. 
The operators $\psi_k$ are defined as~\cite{zimmer:08}
\begin{align}
\psi_k = A_k \cos\theta  - X_{12}^k \sin\theta, 
\end{align}
where $\sin \theta =  \sqrt{N} g_1/\Omega_0 $, 
  $\cos \theta =  \sqrt{2} \Omega_c/\Omega_0$, 
and 
$\Omega_0 = \sqrt{N g_1^2 + 2 \Omega_c^2 }$. 
The operator $A_k
=(a_{k_c + k}   + a_{-k_c + k})/\sqrt{2}$ 
is a superposition of two counterpropagating probe field modes, and $X_{12}^k$ describes the spin coherence, 
$X_{12}^k  = \frac{1}{\sqrt{N}}\sum_{\mu=1}^{N} \Si{12}{\mu} e^{-i k z_{\mu}}$. 
Note that the wave number  $k$ can be positive or negative, and for all 
relevant $k$ we have $|k|\ll k_c$. 
We assume that initially all atoms are in state $\ket{1}$ and that the total number 
of photons is much smaller than the number of atoms $N$. In this case, the dynamics 
induced by the Hamiltonian $H$ is confined to a subspace $\hf$ of the total state space   
where $\bra{\psi}\sum_{\mu=1}^{N} A_{11}^{(\mu)}\ket{\psi}\approx N$ 
for  all $\ket{\psi}\in \hf$. 
It follows that the operators $\psi_k$ obey bosonic commutation relations in  $\hf$, 
$[\psi_k,\psi_p^{\dagger}] = \delta_{kp}$, where we neglected corrections of order $1/N$. 

The dark-state polaritons are eigenstates of $H_{\Lambda}$, but the remaining 
parts $H_0$ and $H_{\text{NL}}$ of the system Hamiltonian give rise to 
a non-trivial time evolution of $\rd$. Fortunately, 
this dynamics can be studied entirely in terms of bosonic quasi-particle excitations 
if the system dynamics is restricted to the subspace $\hf$. 
In particular, the free time evolution $H_0$ introduces a coupling of 
dark-state polaritons to 
bright polaritons, $\phi_k=A_k \sin\theta + X_{12}^k \cos\theta$, and 
photons, $D_k=(a_{k_c + k} - a_{-k_c + k} )/\sqrt{2}$. 
These  excitations  are in turn coupled to the excited state $\ket{3}$. 
 Furthermore,  $H_{\text{NL}}$ introduces a direct coupling of 
dark-state polaritons $\psi_k$  to the excited 
state $\ket{4}$ via a two-particle process~\cite{HY98}. 
Excitations in the  states $\ket{3}$ and $\ket{4}$ 
are created by  $P_{k,+}^{\dagger}$, $P_{k,-}^{\dagger}$ and 
$U_{k,+}^{\dagger}$, $U_{k,-}^{\dagger}$, respectively, 
where 
$ P_{k,\pm}^{\dagger} =  \sum_{\mu = 1}^{N} [ \Si{31}{\mu} e^{i ( k_c +k)z_{\mu}} 
\pm \Si{31}{\mu} e^{-i ( k_c - k)z_{\mu}}]/\sqrt{2N}$ , 
$U_{k,\pm}^{\dagger} =  \sum_{\mu = 1}^{N} [\Si{41}{\mu} e^{i ( k_c +k)z_{\mu}} 
\pm \Si{41}{\mu} e^{-i ( k_c - k)z_{\mu}} ]/\sqrt{2N}$ .
Finally, we note that spontaneous emission from states $\ket{3}$ and $\ket{4}$ results 
in the decay of excitations created by $P_{k,\pm}^{\dagger}$ and $U_{k,\pm}^{\dagger}$.

We employ projection operator techniques~\cite{breuer:os} to derive a master equation for 
 $\rd$ which is obtained from $\vro$ by a partial trace over 
all excitations except for the dark state polaritons $\psi_k$. 
We restrict our analysis to the so-called slow-light regime where $\sin^2\theta\approx 1$ 
and $\cos^2\theta \ll 1$. In this case, the coupling of dark-state polaritons to 
excitations in state $\ket{3}$ and $\ket{4}$ is much slower than the decay of the 
relevant   correlation functions
$\langle\phi_k\phi_p^{\dagger}\rangle(\tau)$, 
$\langle D_k D_p^{\dagger}\rangle(\tau)$, 
and $\langle U_{k,+} U_{p,+}^{\dagger}\rangle(\tau)$, 
which happens on a timescale given  by  the lifetimes 
of the exited states $\ket{3}$ and $\ket{4}$.  
This existence of two different time scales allows us to 
derive a master equation in Born-Markov approximation if
$4 g_2^2\cos^2\theta N_{\text{ph}}\ll \gamma_{42}^2$, 
$\cos^2\theta c^2 k_{\text{max}}^2/\Omega_0^2\ll 1$,
$\cos^2\theta \Delta\omega^2/\Omega_0^2\ll 1$,
and $\Omega_0 \gg \gamma_{ij},|\delta|$. 
Here $c$ is the speed of light, $\Delta\omega=\omega_p-\omega_c$ is the frequency difference between the 
probe  and control fields and $N_{\text{ph}}$ is the number of photons in the pulse. 
We describe the polariton pulse by the field operator 
$\psi(z) = (1/\sqrt{L})\sum_k e^{i k z} \psi_k$ which obeys the  commutation relations 
$[\psi(z),\psi^{\dagger}(z^{\prime})]=\delta(z-z^{\prime})$. 
The  maximal wave number contributing to $\psi$ is  $k_{\text{max}}$. 
Furthermore, our derivation assumes that fast oscillating spin coherences 
with wave number $\pm 2 k_c$ are washed out due to atomic motion~\cite{bajcsy:03}. 
For a small two-photon detuning $\ve = -\cos^2\theta \Delta\omega$, we obtain  
\be 
\hbar \dot{\rd} = -i H_{\text{eff}}\rd +i \rd H_{\text{eff}}^{\dagger} + \mc{I}\rd 
+ \mc{L}_1\rd + \mc{L}_2\rd, 
\label{meq}
\ee
where $H_{\text{eff}}$ is a non-hermitian Hamiltonian, 
\be
H_{\text{eff}} = \frac{\hbar^2}{2 m_{\text{eff}}} \int_0^L \text{d}z
 \partial_z\psi^{\dagger}  \partial_z\psi  
+  \frac{\tilde{g}}{2}\int_0^L \text{d}z  \psi^{\dagger 2}  \psi^2\,,
\label{Heff}
\ee
 $m_{\text{eff}}=-\hbar\Omega_0^2/(2\delta c^2 \cos^2\theta)$ 
is the effective mass of the polaritons, 
$\tilde{g}=2\hbar L g_2^2\cos^2\theta/(\Delta-\cos^2\theta \Delta\omega + i \gamma_{42}/2)$ is 
the complex coupling constant, and 
\begin{align}
& \mc{I}\rd = - \text{Im}(\tilde{g}) \int_0^L \text{d}z \psi^2 \rd  \psi^{\dagger 2}\,,
\label{ird} \\
& \mc{L}_1\rd=   -\frac{\hbar \Gamma \Delta\omega^2  \mc{D}[\psi]}{2\Omega_0^2/\cos^2\theta}  
,  \quad 
\mc{L}_2\rd =  -\frac{\hbar\Gamma   c^2 \mc{D}[\partial_z\psi]}{2\Omega_0^2/\cos^2\theta} .
\end{align}
Here $\mc{D}[\hat{X}]=\int_0^L \text{d}z
(\hat{X}^{\dagger}\hat{X}\rd+\rd\hat{X}^{\dagger} \hat{X}-2\hat{X}\rd\hat{X}^{\dagger})$ 
is a dissipator in Lindblad form~\cite{breuer:os} for an operator $\hat{X}$,  
and $\Gamma=\gamma_{31}+\gamma_{32}$ is the full decay rate of state $\ket{3}$.  
For optical fibers, photon losses  due to leakage are very low and can be neglected. If
they need to be taken into account, an additional decay term with 
the same structure as $\mc{L}_2\rd$ but with a decay rate $\kappa \cos^2\theta$ appears 
($\kappa$ is the bare photon leakage rate). 
To confirm the accuracy of our results, we compared 
the predictions of the master  equation~(\ref{meq}) for the $\Lambda$-
subsystem ($\tilde{g}=0$) to the results of 
a full numerical integration of Maxwell-Bloch equations for 
classical fields and found excellent agreement. 
Next we derive the essential results of this letter from the master 
equation~(\ref{meq}) that describes a one-dimensional system of interacting bosons. 
The first contribution to $H_{\text{eff}}$ in Eq.~(\ref{Heff}), 
$ (\hbar^2/2 m_{\text{eff}}) \int_0^L \text{d}z
 \partial_z\psi^{\dagger}  \partial_z\psi$, 
represents a  kinetic energy term with quadratic dispersion relation for  the polaritons. 
The  term  proportional to $\tilde{g}$ in Eq.~(\ref{Heff}) and 
$\mc{I}\rd$ in Eq.~(\ref{ird}) account for elastic and inelastic two-particle interactions that 
originate from   the coupling of dark-state polaritons to the 
excited state $\ket{4}$. More precisely, the real part of $\tilde{g}$ gives 
rise to a hermitian contribution to $H_{\text{eff}}$ that accounts for 
elastic two-particle collisions. On the other hand, the imaginary part of $\tilde{g}$ 
together with $\mc{I}\rd$ gives rise to a two-particle loss term that can be written 
in Lindblad form as  $-\text{Im}(\tilde{g}/2)\mc{D}[\psi^2]$. 

The contributions   $\mc{L}_1\rd$ and $\mc{L}_2\rd$ describe single-polariton losses 
that can be omitted under the following conditions.  
Since $\mc{L}_1\rd$ is proportional to $\Delta\omega^2$, single-particle losses are minimized by minimizing
$|\Delta\omega|$. Note that this  fact  has not been pointed out so far. From now on we assume that 
$\Delta\omega^2$ is small enough such that $\mc{L}_2\rd$ represents the dominant single particle losses. 
This is reasonable if $|\Delta\omega|$ is at most of the order of GHz 
and implies $|\ve| \ll |\gamma_{24}|$. The term $\mc{L}_2\rd$  is negligible if 
two conditions are met. First, the dynamics induced by the kinetic energy term 
proportional to $m_{\text{eff}}$ in Eq.~(\ref{Heff}) must 
be fast as compared to the inverse decay rate of polaritons introduced by $\mc{L}_2\rd$. This can  
be achieved if we set $|\delta|\gg \Gamma$. 
Second, losses due to $\mc{L}_2\rd$ must be negligible which imposes 
a limit on the maximal evolution time 
$t_{\text{max}}\ll 2\Omega_0^2/(\Gamma c^2 k_{\text{max}}^2\cos^2\theta)$. This implies 
that $t_{\text{max}}$ can be of the order of $1/(\cos^2\theta \Gamma)\gg 1/\Gamma$. 

Under these conditions, the master equation~(\ref{meq}) reduces to 
$\hbar \dot{\rd} = -i H_{\text{eff}}\rd +i \rd H_{\text{eff}}^{\dagger} + \mc{I}\rd$
and  can be identified with  
the generalized Lieb-Liniger model~\cite{duerr:09} for a  one-dimensional 
system of bosons with  mass $m_{\text{eff}}$ and complex interaction parameter $\tilde{g}$. 
All features of the Lieb-Liniger model~\cite{lieb:63,duerr:09}   
are characterized by a single, dimensionless parameter 
$G = m_{\text{eff}} \tilde{g}/(\hbar^2 N_{\text{ph}}/L)$, where $N_{\text{ph}}$ 
is the number of photons in the pulse. The absolute value of $G$ is 
\be
|G| = \frac{g_1^2 g_2^2 L^2 N}{c^2|\delta|\sqrt{\Delta^2 + \frac{\gamma_{42}^2}{4}}N_{\text{ph}}}
= \frac{(1/16) \Gamma \gamma_{42} \text{OD}^2 }{|\delta|
\sqrt{\Delta^2 +\frac{\gamma_{42}^2}{4}} N N_{\text{ph}}}, 
\label{gll}
\ee
where $\text{OD}= 4 N g_1^2 L/(c\Gamma)= 4 N g_2^2 L/(c\gamma_{42})$ is the optical depth 
on the probe field transitions. Note that the parameters $g_{1}^2 L$ and $g_{2}^2 L$ are  
independent of the length of the system since $g_{1},g_{2}\sim 1/\sqrt{A L}$. It follows that the 
parameter $G$ and the optical depth depend only on the transverse area $A$ of the interaction volume, 
but not on the length $L$ of the cell. The absolute value of $G$ characterizes the 
effective interaction strength between the particles. 
In the strongly correlated regime $|G|\gg 1$, the interaction between the particles creates 
a Tonks-Girardeau gas where photons behave like impenetrable hard-core particles that 
never occupy the same position. Formally, this result can be derived via the  pair correlation 
function
$g^{(2)}(z,z^{\prime})=\langle\psi^{\dagger}(z)\psi^{\dagger}(z^{\prime})\psi(z)\psi(z^{\prime})\rangle
/(\langle \hat{n}(z)\rangle \langle \hat{n}(z^{\prime})\rangle)$ 
with $\hat{n}(z)=\psi^{\dagger}(z)\psi(z)$. 
For the ground state of the generalized Lieb-Liniger model in the strongly correlated regime, 
 $g^{(2)}(z,z)= (1-1/N_{\text{ph}}^2)4\pi^2/(3 |G|^2)$ is close to zero and 
vanishes in the limit  $|G|\rightarrow \infty$~\cite{duerr:09}. 
 Moreover, this  ground state  is the same~\cite{duerr:09} as in the 
original model with repulsive interaction for $|G|\rightarrow \infty$. 
It follows that $g^{(2)}(z,z^{\prime})$ for $z\not=z^{\prime}$ exhibits 
Friedel oscillations~\cite{F58} that indicate a crystallization of photons in the fiber.

The parameter $|G|$ is maximal if 
the interaction between the polaritons is purely dissipative ($\Delta = 0$). 
Since the realization of a regime where the  two-particle interactions are dominated 
by elastic processes requires $\Delta\gg\gamma_{42}/2$, the conservative nonlinearities 
are at least an order of magnitude smaller than the dissipative counterparts for $\Delta=0$. 
It follows that purely dissipative interactions between the polaritons we discuss here  
are most effective for the generation of correlations. 

An  analysis of dissipation-induced correlations ($\Delta=0$) requires at least two photons.
Assuming $|\delta|/\Gamma=10$ such that the single-particle loss term 
$\mc{L}_1\rd$ in Eq.~(\ref{meq}) is negligible  and $N_{\text{ph}}=2$,
Eq.~(\ref{gll}) shows that $|G|$ is larger than unity for 
$\text{OD}^2/N > 160$.  A recent experiment~\cite{bajcsy:09} with 
atoms loaded into a hollow fiber reports a value of $\text{OD}^2/N\approx 0.3$.  
If we assume for simplicity that the decay rates of the atomic states $\ket{3}$ 
and $\ket{4}$ do not depend on the transverse area $A$, we find 
$|G|\propto N \lambda_p^4/A^2$, where $\lambda_p$ is the wavelength of 
the probe field.   It follows that $|G|$  does not dependent on the strength of the atomic 
transition dipole moments and could be increased by a reduction of 
the area $A$ or by an increased number of atoms $N$ inside 
the fiber.  In contrast to cavity QED systems~\cite{HP07}, we point out that the 
condition $g_{2} \gg \gamma_{42}$
is not required to obtain 
large values of $|G|$.

The observation of the dissipation-induced  TG gas regime requires 
that the system can be prepared in low-energy states. 
One possibility is  the procedure described in~\cite{chang:08} which 
relies on an adiabatic state transfer realized by a time-dependent detuning $\Delta$. 
A second possibility does not require any tuning of the two-particle losses. 
The master equation~(\ref{meq}) implies that losses due to inelastic two-particle 
interactions are related to the pair-correlation function $g^{(2)}(z,z)$ via 
$\partial_t\langle \hat{n}(z)\rangle = (2/\hbar) \text{Im}(\tilde{g})
 g^{(2)}(z,z) \langle \hat{n}(z)\rangle^2$.  
It follows that uncorrelated states with $g^{(2)}(z,z)\approx 1$ decay much 
faster than those where $g^{(2)}(z,z)\approx 0$.  
Therefore, a regime where $g^{(2)}(z,z)<1$ should be entered on a time scale 
$\hbar/[2\text{Im}(\tilde{g})N_{\text{ph}}/L]$, which is shorter than 
the maximally allowed evolution time $t_{\text{max}}$ if 
$16 g_2^2 N_{\text{ph}}\Omega_0^2/(\Gamma \gamma_{42} c^2k_{\text{max}}^2) > 1$.  
Since the ground state of the generalized 
Lieb-Liniger model decays at the smallest rate~\cite{duerr:09},  
two-particle losses themselves 
are then able to drive the system into states  close to the ground state. 
A more rigorous investigation of this point would require a numerical 
integration of the master equation~(\ref{meq}) which is beyond the scope of this work.

For measurements, we note that the polariton pulse can be released from 
the fiber without distortion if one control field is adiabatically  
switched off~\cite{bajcsy:03,chang:08}. It follows that spatial correlations 
$\langle\psi^{\dagger}(z)\psi(z^{\prime})\rangle$ and
$\langle\psi^{\dagger}(z)\psi^{\dagger}(z^{\prime})\psi(z)\psi(z^{\prime})\rangle$ 
of the trapped pulse are mapped into first and second order correlations in time of the output 
light, respectively. 
Since the latter can be detected via standard quantum optical techniques, 
Friedel oscillations of $g^{(2)}(z,z^{\prime})$, the correlations 
 $\langle\psi^{\dagger}(z)\psi(z^{\prime})\rangle$
and the characteristic momentum distribution of the TG gas, can be measured with high precision.
\begin{acknowledgments}
The authors thank W. Zwerger for discussions. 
This work is part of the Emmy Noether project 
HA 5593/1-1 funded by the German Research Foundation (DFG).
\end{acknowledgments}

\end{document}